\documentstyle[preprint,epsfig,eqsecnum,aps,floats,tighten]{revtex}

\def\met{\mbox{${\hbox{$E$\kern-0.6em\lower-.1ex\hbox{/}}}_T~$}} 
\def\D0{D\O}                            
\def\vs{{\sl vs.}}                      

\newcommand{\Dzero}{D\O\ }

\begin{document}

%
%
\title{Subjet Multiplicity in Quark and Gluon Jets at \Dzero}
\author{\centerline{The \Dzero Collaboration
  \thanks{Submitted to the {\it International Europhysics Conference
        on High Energy Physics},
	\hfill\break
	July 12-18, 2001, Budapest, Hungary,
        \hfill\break 
	and  {\it XX International Symposium on Lepton and Photon Interactions at High Energies}
	\hfill\break
        July 23 -- 28, 2001, Rome, Italy. 
	 }}}
\address{
\centerline{Fermi National Accelerator Laboratory, Batavia, Illinois 60510}
}
%
%
\date{\today}

\maketitle

%
%
\begin{abstract}
We measure the subjet multiplicity $M$ in jets reconstructed with 
a successive combination type of jet algorithm ($k_{\perp}$). 
We select jets with $55<p_{T}<100$ GeV and $\left| \eta
\right| <0.5$. We compare similar samples of jets at $\sqrt{s}=1800$ and
630 GeV. 
The {\small HERWIG} Monte Carlo simulation predicts that 
59\% of the jets are gluon jets at $\sqrt{s}=1800$ GeV, 
and 33\% at $\sqrt{s}=630$ GeV.
Using this information,
we extract the subjet multiplicity in quark ($M_{q}$) and gluon 
($M_{g}$) jets. We also
measure the ratio $R\equiv \frac{\left\langle M_{g}\right\rangle -1}{%
\left\langle M_{q}\right\rangle -1}=
1.84 \pm 0.15{\rm(stat)} ^{+0.22}_{-0.18} {\rm(sys)}$.
\end{abstract}

\newpage
\begin{center}
\end{center}

\normalsize

\vfill\eject

\section{Introduction}
\label{sec:intro}

The Tevatron proton-antiproton collider is a rich environment for studying
high energy physics. The dominant process is jet production, 
described in Quantum Chromodynamics (QCD) by scattering of
the elementary quark and gluon constituents of the incoming hadron beams.
In leading order (LO) QCD, there are two partons in the initial
and final states of the elementary process.
A jet is associated with the energy and momentum of each final state
parton. Experimentally, however, a jet is a cluster of energy
in the calorimeter. Understanding jet structure
is the motivation for the present analysis. 
QCD predicts that gluons radiate more than quarks.  
Asymptotically, the ratio of objects within gluon jets
to quark jets is expected to be in the ratio of their color charges $C_A / C_F = 9 / 4$\cite{Ellis}.

\section{The $\lowercase{k}_T$ Jet Algorithm}
We define jets in the D\O\ detector \cite{D0detector}
with the $k_{\perp}$ algorithm \cite{ES,cat93,cat92}.
The jet algorithm starts with a list of energy
preclusters, formed from calorimeter cells or from particles in
a Monte Carlo event generator.
The preclusters are separated
by $\Delta{\cal R} = \sqrt{ \Delta \eta^2 + \Delta \phi^2} > 0.2$,
where $\eta$ and $\phi$ are the pseudorapidity and azimuthal angle of the
preclusters.
The steps of the jet algorithm are:

1. For each object $i$ in the list, define
$d_{ii} = p_{T,i}^2$, where $p_T$ is the energy transverse to the beam.
For each pair $(i,j)$ of objects, also define
$d_{ij} = min(p_{T,i}^2,p_{T,j}^2) \frac{ \Delta{\cal R}_{ij}^2} {D^2}$,
where $D$ is a parameter of the jet algorithm.

2. If the minimum of all possible $d_{ii}$ and $d_{ij}$ is a $d_{ij}$,
then replace objects $i$ and $j$ by their 4-vector sum and go to step~1.
Else, the minimum is a $d_{ii}$ so 
remove object $i$ from the list and define it to be a jet.

3. If any objects are left in the list, go to step~1.

The algorithm produces a list of jets, each separated by $\Delta{\cal R} > D$.
For this analysis, $D = 0.5$.

The subjet multiplicity is a natural observable
of a $k_{\perp}$ jet\cite{sey_sub2,aleph,sey_sub1,opal,cat93,cat92}.
Subjets are defined by rerunning the $k_{\perp}$ algorithm
starting with a list of preclusters in a jet.
Pairs of objects with the smallest $d_{ij}$ are merged successively until
all remaining $d_{ij} > y_{cut} p_T^{2}(jet)$. 
The resolved objects
are called subjets, and the number of subjets within the jet 
is the subjet multiplicity $M$.
The analysis in this article uses a single resolution 
parameter $y_{cut} = 10^{-3}$.

\section{Jet Selection}
In LO QCD, 
the fraction of final state jets which are gluons decreases
with $x \sim p_T / \sqrt{s}$, 
the momentum fraction of initial state partons within the proton.
For fixed $p_T$, the gluon jet fraction decreases when 
$\sqrt{s}$ is decreased from 1800 GeV to 630 GeV.
We define gluon and quark enriched jet samples 
with identical cuts in events at $\sqrt{s} = 1800$ and 630 GeV
to reduce experimental biases and systematic effects.
Of the two highest $p_T$ jets in the event, 
we select jets with $55 < p_T < 100$ GeV and $|\eta| < 0.5$.

\section{Quark and Gluon Subjet Multiplicity}
\label{subs:ext}
There is a simple method to extract a measurement
of quark and gluon jets on a statistical basis,
using the tools described in the previous sections.
$M$ is the subjet multiplicity 
in a mixed sample of quark and gluon jets.
It may be written as a linear combination of subjet multiplicity
in gluon and quark jets:
\begin{equation}
M=fM_{g}+(1-f)M_{q}
\label{eq:m}
\end{equation}
The coefficients are the fractions of gluon and
quark jets in the sample, $f$ and $(1-f)$, respectively. 
Consider Eq. (\ref{eq:m}) for two
similar samples of jets at $\sqrt{s} = 1800$ and 630 GeV,
assuming $M_{g}$ and $M_{q}$
are independent of $\sqrt{s}$.
The solutions are

\begin{equation}
M_{q}=\frac{f^{1800}M^{630}-f^{630}M^{1800}}{f^{1800}-f^{630}}  \label{eq:mq}
\end{equation}

\begin{equation}
M_{g}=\frac{\left( 1-f^{630}\right) M^{1800}-\left( 1-f^{1800}\right) M^{630}%
}{f^{1800}-f^{630}}  \label{eq:mg}
\end{equation}

where $M^{1800}$ and $M^{630}$ are the
experimental measurements in the mixed jet samples at  
$\sqrt{s} = 1800$ and 630 GeV,
and $f^{1800}$ and $f^{630}$ are the gluon jet fractions in the two samples.
The method relies on knowledge of the two gluon jet fractions.

\section{Results}
\label{subs:mc}

\begin{figure}[htb]
\centerline{\psfig{figure=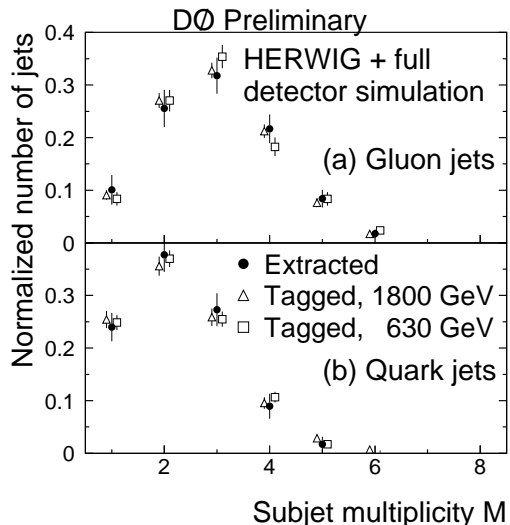,width=7cm}}
\caption{Raw subjet multiplicity in 
fully simulated Monte Carlo quark and gluon
jets. For visibility, we shift the open symbols horizontally.}
\label{fig:mc}
\end{figure}

The {\small HERWIG} 5.9\cite{hw} Monte Carlo event generator
provides an estimate of the gluon jet fractions.
The method is tested using the detector simulation and CTEQ4M PDF.
We tag every selected jet in the detector
as either quark or gluon by the identity of the nearer 
(in $\eta \times \phi$ space)
final state parton in the QCD 2-to-2 hard scatter.
Fig.~\ref{fig:mc} shows that gluon jets in the detector simulation
have more subjets than quark jets.
The tagged subjet multiplicity distributions 
are similar at the two center of mass energies,
verifying the assumptions in \S~\ref{subs:ext}.

We count tagged gluon jets and find
$f^{1800} = 0.59 \pm{0.02}$ and 
$f^{630} = 0.33 \pm{0.03}$,
where the uncertainties are estimated 
from different gluon PDF's.
The nominal gluon jet fractions 
and the Monte Carlo 
measurements at $\sqrt{s} = 1800$ and 630 GeV
are used in Eqs. (\ref{eq:mq}-\ref{eq:mg}).
The extracted quark and gluon jet distributions
in Fig.~\ref{fig:mc} 
agree with the tagged distributions and
demonstrate closure of the method.

\begin{figure}[htb]
\centerline{\psfig{figure=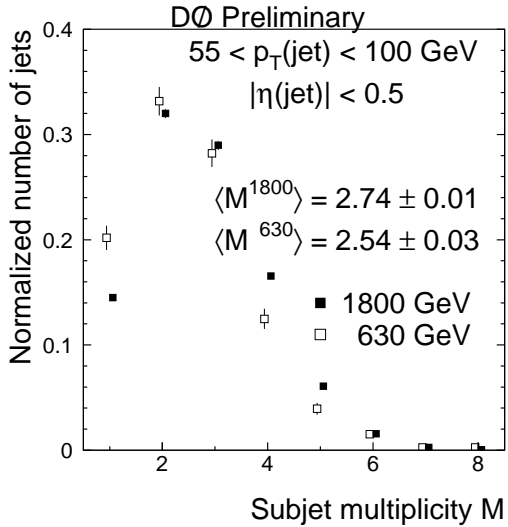,width=7cm}}
\caption{Raw subjet multiplicity in jets from D\O\ data at 
$\sqrt{s} = 1800$ and 630 GeV.}
\label{fig:1800_630}
\end{figure}

Figure~\ref{fig:1800_630} shows the raw subjet multiplicity
in D\O\ data  
at $\sqrt{s} = 1800$ GeV is higher than at $\sqrt{s} = 630$ GeV.
This is consistent with the prediction
that there are more gluon jets at  $\sqrt{s} = 1800$ GeV compared to 
$\sqrt{s} = 630$ GeV, and gluons radiate more than quarks.
The combination of the distributions in Fig.~\ref{fig:1800_630}
and the gluon jet fractions
gives the raw subjet multiplicity
distributions in quark and gluon jets, 
according to Eqs. (\ref{eq:mq}-\ref{eq:mg}). 

\begin{figure}[htb]
\centerline{\psfig{figure=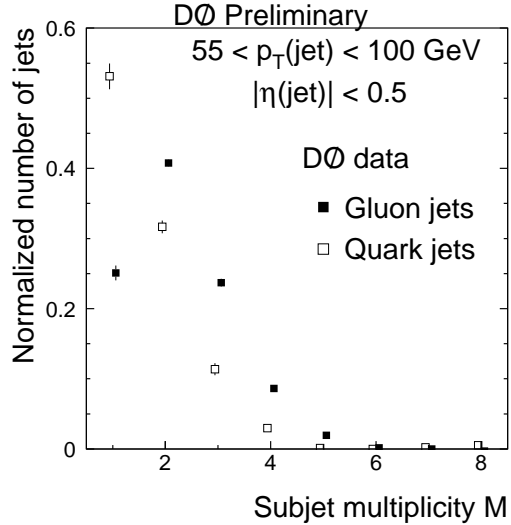,width=7cm}}
\caption{Corrected subjet multiplicity in quark and gluon jets, 
extracted from D\O\ data.}
\label{fig:qg}
\end{figure}

The quark and gluon raw
subjet multiplicity distributions
need separate corrections for various detector-dependent effects. 
These are derived from Monte
Carlo, which describes the raw D\O\ data well.
Each Monte Carlo jet
in the detector simulation is matched (within $\Delta{\cal R} < 0.5$)
to a jet reconstructed from
particles without the detector simulation. 
We tag detector jets as either quark or gluon,
and study the subjet multiplicity in particle jets $M^{ptcl}$
\vs ~that in detector jets $M^{det}$.
The correction unsmears $M^{det}$ to give $M^{ptcl}$,
in bins of $M^{det}$.
Figure~\ref{fig:qg} shows the corrected subjet multiplicity is 
clearly larger for gluon jets
compared to quark jets.

The gluon jet fractions are the largest
source of systematic error. 
We vary the gluon jet fractions by the uncertainties
in an anti-correlated fashion at the two values of $\sqrt{s}$
to measure the effect on $R$.
The systematic errors listed in Table~\ref{tab:sys}
are added
in quadrature to 
obtain the total uncertainty in the corrected ratio
$R = \frac{\langle M_g \rangle - 1} {\langle M_q \rangle - 1}
= 1.84 \pm 0.15{\rm(stat)} ^{+0.22}_{-0.18} {\rm(sys)}$.

\begin{table}[htb]
\begin{center}
\begin{tabular}{lr}
\multicolumn{1}{c} {Source} &
\multicolumn{1}{c} {$\delta R$} \\
\hline\hline
Gluon Jet Fraction    	& $^{+0.17}_{-0.10}$ \\
Jet $p_T$ cut		& $\pm 0.12$ \\
Detector Simulation   	& $\pm 0.02$ \\
Unsmearing          	& $\pm 0.07$ \\
\end{tabular}
\caption{Systematic Errors.}
\label{tab:sys}
\end{center}
\end{table}

\section{Conclusion}
We extract the $y_{cut} = 10^{-3}$ 
subjet multiplicity in quark and gluon jets
from measurements of mixed jet samples at 
$\sqrt{s} = 1800$ and 630 GeV.
On a statistical level,
gluon jets have more subjets than quark jets.
We measure the ratio
of additional subjets in gluon jets to quark jets
$R = 1.84 \pm 0.15{\rm(stat)} ^{+0.22}_{-0.18} {\rm(sys)}$.
The ratio is well described by the {\small HERWIG} parton 
shower Monte Carlo, and is only slightly 
smaller than the naive QCD prediction 9/4.

\section*{Acknowledgements}
\label{sec:ack}
%
%
We thank the staffs at Fermilab and collaborating institutions for
their contributions to this work, and acknowledge support from the
Department of Energy and National Science Foundation (U.S.A.),
Commissariat  \` a L'Energie Atomique (France), State Committee for
Science and Technology and Ministry for Atomic    Energy (Russia),
CAPES and CNPq (Brazil), Departments of Atomic Energy and Science and
Education (India), Colciencias (Colombia), CONACyT (Mexico), Ministry
of Education and KOSEF (Korea), and CONICET and UBACyT (Argentina). 
%

\end{document}